\documentstyle[12pt,epsfig]{article}
\def \ins#1#2#3#4#5#6#7 {
  \begin{figure}[#1]
    \begin{center}
      \psfig{file=#2,width=#3,height=#4,angle=#5}
      \caption{#6}
      \label{#7}
    \end{center}
  \end{figure}
  }
\begin{document}
%\large

\begin{center}
{\Large \bf Parametrization of the
$\bar PP$ Elastic Scattering Differential Cross Section
Between 2~GeV/c~$ \leq P_{lab}\leq$~16~ GeV/c
}
\end{center}

\begin{center}
A. Galoyan\footnote{Joint Institute for Nuclear Research, Dubna, Russia.},
J. Ritman\footnote{Institute for Nuclear Physics, Forschungszentrum J\"ulich,
                   and Ruhr University Bochum, Germany.},
A. Sokolov\footnote{Institute for Nuclear Physics, Forschungszentrum J\"ulich, Germany.},
V. Uzhinsky$^{1}$ \\

\end{center}

\begin{center}
\begin{minipage}{13cm}
A parameterization of the $\bar pp$ differential elastic scattering cross
section in the beam momentum range from 2 to 16 GeV/c is proposed.
The parameterization well describes the existing data including the
observed diffraction pattern at four-momentum transfer $|t|$ up to 1.5-2.0~GeV$^2$.
It can be used for detailed calculations of the
radiation load on the detectors being designed for the PANDA detector at the future FAIR
facility in Darmstadt.
\end{minipage}

\end{center}

\vspace{0.5cm}

\section{Introduction}
\label{sec:intro}
The science goals underlying the international FAIR project \cite{FAIR}
that is being realized in Darmstadt span a broad range of research activities on
the structure of matter. One component of this facility is directed towards
studies of hadronic matter at the sub-nuclear level with beams of antiprotons.
These studies focus on two key aspects: confinement of quarks and the
 generation of the hadron masses. They are intimately related to the
existence (and spontaneous breaking) of chiral symmetry, a fundamental
property of the strong interaction. These goals will be persuaded by performing
precision measurements of charged and neutral decay products from antiproton-proton
annihilation in the charmonium mass region.  These high rate experiments place
demanding requirements on the materials and detectors employed. Since up to one half of
the total antiproton-proton reaction cross section in the beam momentum range of
interest is elastic scattering, a good description of the $\bar pp$ elastic
scattering differential cross section is required in order to quantitatively
assess the detectors being designed. This parameterization is additionally
required since the beam monitoring and luminosity control at the experiment
will be implemented via measuring the elastic $\bar pp$ scattering.

\subsection{The High Energy Storage Ring (HESR) for antiprotons}
\label{ssec:HESR}
The future Facility for Antiproton and Ion Research -
FAIR - in Darmstadt will include a storage ring for beams of phase space cooled antiprotons
with unprecedented quality and intensity. The antiproton beam will be produced
by a primary proton beam from the planned fast cycling, superconducting 100 T-m ring. The
antiprotons will be collected with an average rate of about 10$^7$/s and then stochastically
cooled and stored. After $5\times 10^{10}$ antiprotons have been produced, they will be
transferred to the HESR where internal experiments in the beam momentum range 1.5 -- 15 GeV/c
can be performed.

The HESR is designed as a racetrack shaped storage ring with a maximum magnetic bending
power of 50 Tm. The storage ring will have a circumference of 574 m, including two 132 m
long straight sections. One of these sections will be mainly used for the installation
of an internal target in combination with a large detector system. The opposite long
straight section is used for beam injection, beam acceleration and electron cooling.
In addition a stochastic cooling system for transverse and longitudinal cooling is foreseen
to be installed at the entrance and exit of straight sections. The latter has to be
designed to allow for experiments with either high momentum resolution of about $10^{-5}$
at reduced luminosity or at high luminosity up to $2 \times 10^{32} /cm/s$ with enlarged
momentum spread.

\subsection{Hadron Spectroscopy with Antiproton Annhilation at the PANDA detector}
\label{ssec:PANDA}
The PANDA experiment, located at an internal target position of the High Energy Storage Ring
for anti-protons, is one of the large installations at
the future FAIR facility~\cite{PANDA}. It is being planned by a multi-national collaboration,
currently consisting of about 350 physicists from 50 institutions in 15 countries. The PANDA
detector is designed as a multi-purpose setup.  The cornerstones
 of the PANDA physics program are:
\begin{itemize}
\item Study of narrow charmonium states at so far unprecedented precision;
\item Search for gluonic excitations such as hybrids and glueballs in the charmonium mass region;
\item Investigate the properties of mesons with hidden and open charm in the nuclear medium;
\item Spectroscopy of double strange hypernuclei.
\end{itemize}
The experiment will use internal targets. It is conceived to use either pellets of frozen
$H_2$ or cluster jet targets for the $\bar pp$ reactions, and wire targets for the $\bar pA$
reactions. Pellet targets such as the one in operation at WASA at COSY 
shoot droplets of frozen $H_2$ with radii 20-40 $\mu$m with typical separations of
1 mm transversely through the beam.

This detector facility must be able to handle high rates (10$^7$ annihilations/s ),
with good particle identification and momentum resolution for $\gamma,~ e,~ \mu,~ \pi,~ K$,
and $p$. Furthermore, the detector must have the ability to measure $D,~ K^0_S$, and $\Lambda$
which decay at displaced vertices. Finally, a large solid angle coverage is essential for
partial wave analysis of resonance states.

In order to cope with the variety of final states and the large range of particle momenta
and emission angles, associated with the different physics topics, the detector has
almost 4$\pi$ detection capability both for charged particles and photons. It is divided
into two sub-components, a central target spectrometer and a
forward spectrometer with an overall length of 12 m of the total detector.

The beam monitoring and control at the experiment will be implemented via measurement of
elastic $\bar pp$ scattering. Thus, it is very important to have a good systematic
description of known experimental data. It is a subject of the paper. In Sec.~1 we give
the main formulae. The fitting procedure is described in Sec.~2. A short conclusion is presented
at the end of the paper.

\section{$\bar PP$ Elastic Scattering}
\label{sec:1}
A collection of the differential cross section 
data for $\bar pp$ elastic
scattering with beam momentum above 1 GeV/c  is
presented in Fig.~1 \cite{1_077}--\cite{25_2_A_40_1}.
\ins{cbth}{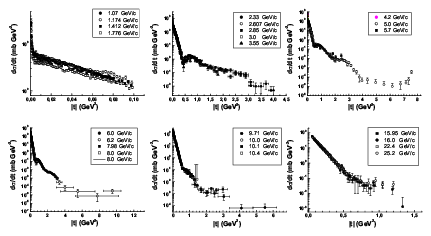}{160mm}{100mm}{0}
{Differential elastic cross sections of $\bar pp$
scattering at various energies. Points are experimental data from Durham HEP Databases.}
\noindent
As seen from the figure, at low beam momenta (1 -- 2 GeV/c) Coulomb scattering
dominates at low 4-momentum transfer ($|t|<~ 0.05~$ GeV$^2$).
At higher energies a dip appears in the region $|t|\sim 0.4$ GeV$^2$.
Above $P_{lab}> 4$ GeV/c an additional diffractional dip appears near $|t|\sim 2$ GeV$^2$.

At low momentum transfer the cross section is usually parameterized as
$$
\frac{d\sigma}{dt}=\frac{\pi}{k^2}\left| f_C e^{i\delta}+f_H\right|^2=
\frac{d\sigma_C}{dt}+\frac{d\sigma_{int}}{dt}+\frac{d\sigma_H}{dt},
$$
where
\begin{eqnarray}
\frac{d\sigma_C}{dt}&=&\frac{4\pi \alpha^2_{EM} G^4(t)}{\beta^2t^2};\\ \nonumber
\frac{d\sigma_{int}}{dt}&=&\frac{\alpha_{EM}\sigma_{Total}}{\beta |t|} G^2(t)e^{\frac{1}{2}Bt}
(\rho cos\delta + sin\delta) \\ \nonumber
\frac{d\sigma_H}{dt}&=&\frac{\sigma^2_{Total} (1+\rho^2)}{16 \pi} e^{Bt}. \nonumber
\end{eqnarray}
Here, $d\sigma_C/dt$ and $d\sigma_H/dt$ are the Coulomb and hadronic parts of the cross section,
respectively. $d\sigma_{int}/dt$ represents the interference term. $\alpha_{EM}$ is the fine
structure constant.

The proton dipole form factor $G(t)=(1+\Delta)^{-2}$, where
$\Delta=|t|/0.71$. The Coulomb phase
$$
\delta(t)=\alpha_{EM}\left[ 0.577+ln\left(\frac{B|t|}{2}-4 \Delta\right)+ 4\Delta ln(4\Delta) + 2\Delta\right].
$$
$\sigma_{Total}$ is the total hadronic cross section, and $B$ is the so-called slope parameter.
$\rho$ is the ratio of real to imaginary parts of the hadronic scattering amplitude at zero
momentum transfer.
The hadronic part of the amplitude can be parameterized as
a simple exponent, $f_H \propto e^{\frac{1}{2}Bt}.$

Neglecting $\rho$ and integrating $d\sigma_H/dt$, one finds
$B$ to be
$$
B=\frac{\sigma^2_{Total}}{16 \pi \sigma_{elastic}}.
$$
Using the Particle Data Group's parameterization for the total and elastic $\bar pp$ scattering
cross section \cite{PDG96}, it is easy to calculate $B$ at a given energy.
$$
\sigma_{Total}=38.4+77.6 P_{lab}^{-0.64}+0.260 ln^2(P_{lab})-1.20 ln(P_{lab}),~ (mb),
v$$
$$
\sigma_{Elastic}=10.2+52.7 P_{lab}^{-1.16}+0.125 ln^2(P_{lab})-1.28 ln(P_{lab}),~ (mb).
$$
This simple parameterization is indicated below by the
solid line in Figs~3--5.
As seen in Figs.~3--5, such parameterization can be applied at $P_{lab}\geq 1.1$
GeV/c and $|t|\leq 0.2~-~0.3$ GeV$^2$.

In order to describe the differential cross section in a wider range
of $t$, a $\chi^2$ minimization of the following expression has been applied to the data:
$$
\frac{d\sigma}{dt}=A_1\cdot \left[ e^{t/2t_1}-A_2\cdot e^{t/2t_2}\right]^2 +A_3\cdot e^{t/t_2}.
$$

{\small
\begin{table*}
\centering
\caption {Fit parameters to the measured data at the given beam momenta.}
\label{tab:1}
\begin{tabular}{|r|r|r|r|r|r|r|r|} \hline
$P_{lab}$ &   $A_1$          &         $t_1$       &       $A_2$       &        $t_2$      &   $A_3$          &$\chi^2$& N  \\ \hline
 2.33     & 528.4 $\pm$ 14.0 & 0.085 $\pm$ 0.002   & 0.137 $\pm$ 0.014 & 0.430 $\pm$ 0.028 &  2.21 $\pm$ 0.27 & ~~48.6 & 66 \\
 2.85     & 443.1 $\pm$ 13.3 & 0.094 $\pm$ 0.002   & 0.172 $\pm$ 0.016 & 0.377 $\pm$ 0.020 &  1.82 $\pm$ 0.21 & ~~74.9 & 88 \\
 5.00     & 268.0 $\pm$ 15.7 & 0.113 $\pm$ 0.003   & 0.350 $\pm$ 0.033 & 0.221 $\pm$ 0.006 &  2.11 $\pm$ 0.12 & 250.8& 86   \\
 5.70     & 237.9 $\pm$ 11.4 & 0.091 $\pm$ 0.003   & 0.106 $\pm$ 0.026 & 0.366 $\pm$ 0.409 &  0.92 $\pm$ 0.23 & ~~21.9 & 47 \\
 6.20     & 279.1 $\pm$ 11.0 & 0.086 $\pm$ 0.002   & 0.136 $\pm$ 0.009 & 0.264 $\pm$ 0.004 &  1.00 $\pm$ 0.02 & 339.2& 70   \\
10.10     & 276.9 $\pm$ 41.1 & 0.096 $\pm$ 0.006   & 0.218 $\pm$ 0.080 & 0.193 $\pm$ 0.018 &  3.11 $\pm$ 0.74 & ~~50.8 & 35 \\
10.40     & 173.2 $\pm$ ~~6.9  & 0.090 $\pm$ 0.001 & 0.079 $\pm$ 0.020 & 0.284 $\pm$ 0.033 &  0.97 $\pm$ 0.30 & ~~42.9 & 61 \\
15.95     & 140.3 $\pm$ 54.7 & 0.099 $\pm$ 0.010   & 0.171 $\pm$ 0.182 & 0.203 $\pm$ 0.065 &  2.13 $\pm$ 2.27 & ~~~7.4  & 23\\
16.00     & 108.4 $\pm$ ~~6.5  & 0.094 $\pm$ 0.004 & 0.034 $\pm$ 0.039 & 0.615 $\pm$ 0.683 &  0.17 $\pm$ 0.26 & ~~~8.6 & 34 \\\hline
\end{tabular}
\end{table*}
}

 The parameters are presented in Table \ref{tab:1} and Fig.~2.
This parameterization well describes most  part
 of the data, with the following notable exceptions. The data at $P_{lab}=3.55$ and $3.66$
GeV/c were not included in the table \ref{tab:1}, because there were only few
points at small $|t|$.
The data at $P_{lab}=6.2$ GeV/c \cite{6_2} gave the large value of $\chi^2$
due to the same reason. The fit of the data at
$P_{lab}=16$ GeV/c \cite{8_A_16} resulted in too small value for $A_3$, because the points at large
$t$ were not presented. The situation with the data at $P_{lab}=5$ and
$10.1$ GeV/c \cite{5_0, 10_1} was not so clear.

\begin{figure}[cbth]
  \begin{center}
    \psfig{file=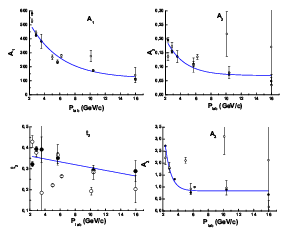,width=160mm,height=120mm,angle=0}
    \caption{Energy dependencies of the fitted parameters. The light points are the
             data of the tabl.~1. Dark points are from the tabl.~2.}
    \label{fig2}
    \psfig{file=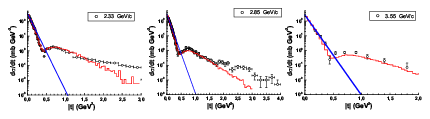,width=160mm,height=80mm,angle=0}
%    \vspace{-15mm}
    \caption{Differential elastic cross sections of $\bar pp$
scattering at $P_{lab}=2-4$ GeV/c. Points are experimental data
\protect{\cite{2_33,2_85,3_55}}. Dashed lines are results of single exponential
paramereizations. Histograms represent our parametrization.}
    \label{fig3}
  \end{center}
\end{figure}

\begin{figure}[cbth]
  \begin{center}
    \psfig{file=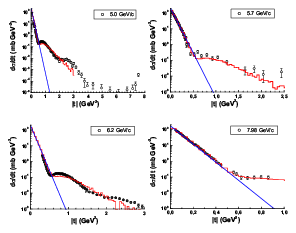,width=160mm,height=100mm,angle=0}
    \caption{The same as in Fig.~3 for $P_{lab}=4-8$ GeV/c. Points are experimental
             data \protect{\cite{5_0,5_7,6_2,7_97_A_15_95}}.}
    \label{fig4}
    \psfig{file=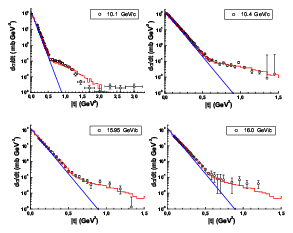,width=160mm,height=100mm,angle=0}
%    \vspace{-15mm}
    \caption{The same as in Fig.~3 for $P_{lab}=8-16$ GeV/c. Points are experimental
             data \protect{\cite{10_1,10_4,7_97_A_15_95,8_A_16}}.}
    \label{fig5}
  \end{center}
\end{figure}

The parameterization did not allow to determine
a regular dependence of the parameters on the beam momentum.
The next step was to redo the minimization of the function to the data,
excluding the data at $P_{lab}=3.66$, $5$, $6.2$, $10.1$, and
$16$ GeV/c. It was assumed that $t_1=0.0899$ (an average value of $t_1$ in
Table \ref{tab:1}) in order to reduce the number of the parameters
The results are presented in Table \ref{tab:2} and Fig.~2.
The energy dependence of the parameters becomes more regular.

\begin{table*}
\centering
\caption {Summary of the parameters of the constrained fit to the differential cross section data.}
\label{tab:2}
\begin{tabular}{|r|r|r|r|r|r|r|} \hline
$P_{lab}$ &   $A_1 $       &   $A_2$           &          $t_2$    &   $A_3$         &$\chi^2$& N   \\ \hline
 2.33     & 582 $\pm$ 17.0 & 0.196 $\pm$ 0.008 & 0.322 $\pm$ 0.012 & 2.72 $\pm$ 0.38 & 57.6   & 66  \\
 2.85     & 426 $\pm$ ~8.5 & 0.153 $\pm$ 0.005 & 0.394 $\pm$ 0.012 & 1.78 $\pm$ 0.19 & 84.9   & 88  \\
 3.55     & 382 $\pm$ 50.2 & 0.137 $\pm$ 0.021 & 0.392 $\pm$ 0.060 & 1.33 $\pm$ 0.42 & 10.7   & 12  \\
 5.7      & 232 $\pm$ ~7.4 & 0.110 $\pm$ 0.010 & 0.351 $\pm$ 0.028 & 0.77 $\pm$ 0.20 & 28.4   & 47  \\
10.40     & 171 $\pm$ ~1.9 & 0.074 $\pm$ 0.004 & 0.293 $\pm$ 0.015 & 0.89 $\pm$ 0.12 & 42.9   & 61  \\
15.95     & 113.$\pm$ ~4.2 & 0.049 $\pm$ 0.013 & 0.289 $\pm$ 0.049 & 0.69 $\pm$ 0.32 & 8.47   & 23  \\\hline
\end{tabular}
\end{table*}

In order to interpolate the parameters presented here to other beam momenta in the
range $2 < P_{Lab} < 16$ GeV/c, the results in Table \ref{tab:2} have been parameterized
as follows:
\begin{center}
\begin{tabular}{l}
$A_1= 115.0+ 650.0\cdot e^{-P_{lab}/4.08},$\\
$t_1=0.0899,$                               \\
$A_2=0.0687+ 0.307\cdot e^{-P_{lab}/2.367},$\\
$t_2=-2.979+ 3.353\cdot e^{-P{lab}/0.67009},$\\
$A_3=0.11959+3.86474 \cdot e^{-P_{lab}/0.765}.$\\
\end{tabular}
\end{center}

This parameterization is indicated by the solid lines in 
Fig.~2.
A description of the experimental data is presented in Figs. 3--5.
As seen, we have a good description of main part of the data in the region of
$t\le$ 1.5 -- 2.0 GeV/c$^2$. However, there is a regular discrepancy between the data
at $P_{lab}=5$ GeV/c \cite{5_0} and the parameterization in the region
of the second maximum. We suppose that the data at $P_{lab}=6.2$
GeV/c \cite{6_2} are distorted too strong, and are not in an agreement
with common regularity. The data at $P_{lab}=10.1$ \cite{10_1} are
reproduced only roughly in the second maximum. It would be well to
re-measure the data at pointed momenta.

For Monte Carlo simulation of the elastic scattering,
$d\sigma/dt/\sigma_{elastic}$ was presented as a sum of two distributions:

$$
\frac{1}{\sigma_{elastic}}\frac{d\sigma}{dt}=d_1(t)+d_2(t),
$$
$$
d_1(t)=A_1\cdot \left[ e^{t/2t_1}-A_2\cdot e^{t/2t_2}\right]^2/\int \frac{d\sigma}{dt}dt',
$$
$$
d_2(t)=A_3\cdot e^{t/t_2}/\int \frac{d\sigma}{dt}dt'.
$$

 Sampling of $t$ according to the second distribution was performed by the formulae
$$
t=t_2 ln\left[ 1-\xi (1 - e^{t_{max}/t2})\right],
$$
where $\xi$ is random number uniformly distributed in the interval [0,1].

The first distribution was re-written as
$$
d_1(t)= \left[ e^{t/t_1} +A_2^2 \cdot e^{t/t_2}\right]
\frac {\left[ e^{t/2t_1}-A_2\cdot e^{t/2t_2}\right]^2}  {\left[ e^{t/t_1}+A_2^2 \cdot e^{t/t_2}\right]}
/\int \frac{d\sigma}{dt}dt'.
$$

The expression in the first brackets was considered as a distribution, and the fraction was
taken as rejection function.

\ins{cbth}{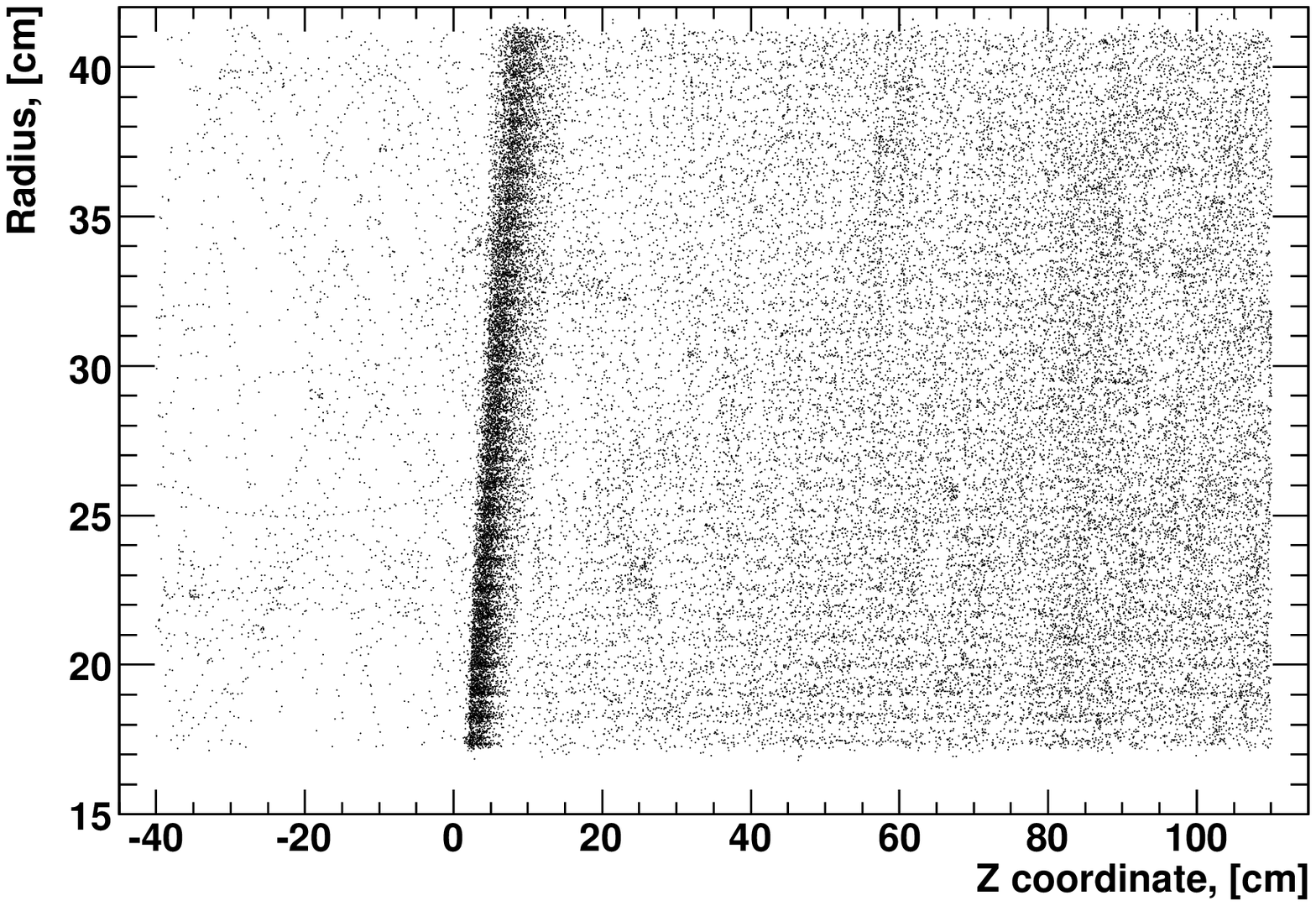}{160mm}{50mm}{0}{Hit positions in the  Straw Tube Tracker detector of
the PANDA experiment  due to the elastic scattering.}{fig6}

Now, the proposed parameterization and the Monte Carlo algorithm
 are implemented in PANDA computational framework
which allows one to estimate an influence of the elastic scattering on the PANDA
sub-detectors.
To understand expected results, a simple consideration can be applied. In PANDA experiment, the target
will be surrounding by beam pipe. Thus, the scattering anti-protons  with $\theta <10^o$,
and the recoil protons with energy $T< 20$ MeV flying with polar angle near to $90^o$ will not
be registered. The differential cross-section falls down very quickly with decreasing
of emission angle of the recoil protons.
So, the recoil protons will be registered by central tracking detector
as a broad jet with main axis depending on the beam energy. Results of direct simulation of hit
positions in Straw Tube Tracker detector due to the elastic scattering
are in agreement with the above given consideration (Fig.~6).
 According to it, the elastic scattering will produce non-uniform radiation
 load in the central part of the detector.

Effect of the elastic scattering has to be taken into account
at design of the PANDA sub-detectors, and for creation of the beam monitor.

\end{document}